\documentclass[aps,amsmath,amssymb,prb,reprint,superscriptaddress]{revtex4-2}
\usepackage{graphicx}

\newcommand{\srtwo}{Sr$_2$RuO$_4$}
\newcommand{\lsco}{La$_{2-x}$Sr$_x$CuO$_4$}
\newcommand{\pdcoo}{PdCoO$_2$}

\newcommand{\salt}{$\kappa$-(BEDT-TTF)$_2$Cu[N(CN)$_2$]Br}

\newcommand{\sab}{$S_{ab}$}
\newcommand{\scc}{$S_{c}$}

\newcommand{\seebeck}[1]{\ensuremath{#1\,\mu\textnormal{V K}^{-1}}}
\newcommand{\sovert}[1]{\ensuremath{#1\,\mu\textnormal{V K}^{-2}}}

\begin{document}
\title{Anisotropic Seebeck coefficient of \srtwo{} in the incoherent regime.}

\author{Ramzy~Daou}
\affiliation{Normandie Univ, ENSICAEN, UNICAEN, CNRS, CRISMAT, 14000 Caen, France}

\author{Sylvie~H\'ebert}
\affiliation{Normandie Univ, ENSICAEN, UNICAEN, CNRS, CRISMAT, 14000 Caen, France}

\author{Ga\"el Grissonnanche}
\affiliation{D\'epartement de physique \& RQMP, Universit\'e de Sherbrooke, Sherbrooke, Qu\'ebec J1K 2R1, Canada}

\author{Elena Hassinger}
\affiliation{D\'epartement de physique \& RQMP, Universit\'e de Sherbrooke, Sherbrooke, Qu\'ebec J1K 2R1, Canada}

\author{Louis Taillefer}
\affiliation{D\'epartement de physique \& RQMP, Universit\'e de Sherbrooke, Sherbrooke, Qu\'ebec J1K 2R1, Canada}

\author{Haruka Taniguchi}
\affiliation{Department of Physics, Kyoto University, Kyoto 606-8502, Japan}
\affiliation{Department of Applied Physics, Graduate School of Engineering, Nagoya University, Nagoya, 464-8603, Japan}

\author{Yoshiteru Maeno}
\affiliation{Department of Physics, Kyoto University, Kyoto 606-8502, Japan}
\affiliation{Toyota Riken – Kyoto Univ. Research Center (TRiKUC), Kyoto 606-8501, Japan}

\author{Alexandra S. Gibbs}
\affiliation{Max Planck Institute for Chemical Physics of Solids, N\"{o}thnitzer Str. 40, 01187 Dresden, Germany}

\author{Andrew P. Mackenzie}
\affiliation{Max Planck Institute for Chemical Physics of Solids, N\"{o}thnitzer Str. 40, 01187 Dresden, Germany}

\date{\today}

\begin{abstract}
Intuitive entropic interpretations of the thermoelectric effect in metals predict an isotropic Seebeck coefficient at high temperatures in the incoherent regime even in anisotropic metals since entropy is not directional. \srtwo{} is an enigmatic material known for a well characterised anisotropic normal state and unconventional superconductivity. Recent ab-initio transport calculations of \srtwo{} that include the effect of strong electronic correlations predicted an enhanced high-temperature anisotropy of the Seebeck coefficient at temperatures above 300 K, but experimental evidence is missing. From measurements on clean \srtwo{} single crystals along both crystallographic directions, we find that the Seebeck coefficient becomes increasingly isotropic upon heating towards room temperature as generally expected. Above 300 K, however, $S$ acquires a new anisotropy which rises up to the highest temperatures measured (750 K), in qualitative agreement with calculations. This is a challenge to entropic interpretations and highlights the lack of an intuitive framework to understand the anisotropy of thermopower at high temperatures.
\end{abstract}
\maketitle


The layered perovskite \srtwo{} has been studied intensively since it was found to host an unconventional superconducting ground state\cite{Mackenzie2017}. A wide range of thermodynamic, transport and spectroscopic experiments have examined both the superconducting \cite{Mackenzie2003} and normal states \cite{Bergemann2003}, especially at low temperatures. \textit{Ab initio} electronic structure calculations provide increasingly good agreement with experimental observations, but consensus is lacking on the precise nature of the superconducting order parameter as well as the pairing mechanism.

The normal state electronic structure is characterised by three bands arising mainly from Ru 4d states. A renormalised ground state with enhanced effective mass indicates the presence of strong correlations. Above the modest Fermi liquid temperature of $\sim$25K, there is a crossover from the regime of coherent, well-realised quasiparticles into the `incoherent' regime where the excitation spectrum lacks a well-defined Drude peak. 
A feature of note is the peak in c-axis resistivity that occurs around 150K that has yet to be satisfactorily explained \cite{Tyler1998}.

It has only become possible for \textit{ab initio} calculations to reproduce experimentally observed trends in transport properties in strongly correlated materials in the last few years. Recently, Dynamical Mean-Field Theory (DMFT) calculations have predicted an enhancement of the out-of-plane Seebeck coefficient of \srtwo{} at high temperatures \cite{Mravlje2016}.

The Seebeck coefficient $S$ quantifies the relationship between charge and heat transport in a conductor. While it is a transport property, it does not depend on the dimensions of the sample. Its existence can be derived in the absence of scattering purely from thermodynamic considerations \cite{Callen1948}. It has therefore been argued that in limit of either low \cite{Behnia2004} or high temperature \cite{Chaikin1976,Peterson2010}, it scales linearly with the entropy associated with the charge carriers. A further implication is that in these limits $S$ should be isotropic, since the entropy is not a directional quantity. This interpretation makes it useful in separating, for example, the entropy associated with charge carriers from other sources such as lattice vibrations.

In this Letter, we show measurements of the Seebeck coefficient of \srtwo{} in both in- and out-of-plane directions on high quality single crystals up to 750\,K that confirm the trends predicted by numerical calculcations. Upon heating up towards room temperature, the Seebeck coefficient becomes increasingly isotropic, as generally expected. Above around 300 K, which corresponds to the regime of incoherent transport, $S$ acquires a new anisotropy, which rises up to the highest temperatures measured (750 K). This high-temperature anisotropy is the subject of this paper. 



The Seebeck coefficient was measured using a standard one-heater, two-thermometer steady-state technique from 2-320\,K as described in Ref.\onlinecite{Daou2015}. Fine-wire thermocouples were attached directly to the samples to measure the temperature difference. The thermoelectric voltage was measured against  phosphor-bronze reference leads. The contribution to the Seebeck signal was negligible ($<\seebeck{0.1}$) in this range and has not been subtracted from the data.

In the temperature range 300-800\,K, $S$ was measured using a two-heater, two-thermometer technique on a miniature high-temperature sample holder designed to fit into a standard PPMS cryostat. The reduced size of the experiment allows us to reach high temperatures without applying more heat than the cryostat can safely dissipate. Two two-wire type-E thermocouples made of 25$\mu$m diameter wire were used to measure the temperature at two points along the sample. Two platinum-wire heaters were used to apply the temperature gradient, which could be reversed. The thermoelectric voltage was measured with reference to the chromel leads of the thermocouples. The temperature and voltage were therefore sensed at exactly the same points, directly at the sample. The Seebeck coefficient was extracted from the linear slope of the voltage as a function of the applied temperature difference between the thermocouples; no non-linearity was observed. To obtain the Seebeck coefficient of the chromel reference leads, we measured a high-purity platinum wire as a reference up to 800\,K and subtracted its contribution \cite{Burkov2001} which is known with an accuracy of around $\pm\seebeck{0.2}$. The precision of the measurement is indicated by the scatter of the points ($<$\seebeck{1} everywhere). If there are additional systematic errors associated with, e.g. thermal radiation losses at high temperatures, the use of very similarly sized samples ensures that the results are comparable.

The \srtwo{} single crystals used in this study were grown by a floating zone technique. For ab-plane measurements we used a crystal of size $3.0\times 0.8\times 0.3$ mm$^3$. For c-axis measurements the sample measured $0.3\times 0.4\times 3.0$ mm$^3$.

\begin{figure}
 \includegraphics[width=0.48\textwidth]{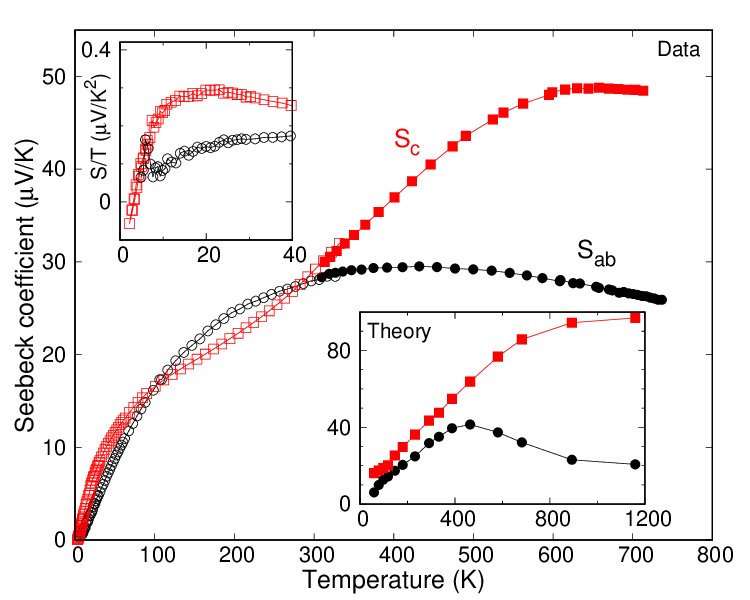}
 \caption{Seebeck coefficient of \srtwo{} up to 800\,K for samples \#1 (\sab{}, circles) and \#2 (\scc{}, squares). The low temperature (open symbols) and high temperature (closed symbols) data overlap well. \scc{} continues to increase above 300\,K while \sab{} reaches a maximum around 400\,K and decreases slightly thereafter. Upper inset: low temperature data plotted as $S/T$. Lower inset: DMFT calculations of \sab{} and \scc{} from Ref.~\onlinecite{Mravlje2016}.}
 \label{fig:thermopower}
\end{figure}

The results over an extended temperature range up to 800\,K are shown in Figure~\ref{fig:thermopower}. The overlap between high- and low-temperature data from different experimental setups is good, which validates the experimental protocols particularly with respect to the use of different reference leads (phosphor-bronze at low temperature; chromel at high temperature).

Up to 300\,K \sab{} matches existing results \cite{Yoshino1996,Xu2008} well. There is a smooth rise from zero with a gradual decrease of slope. \scc{} meanwhile is of similar magntiude to \sab{} below 200\,K, although there are changes in slope where \sab{} appears smooth, and the approach to low temperatures of \scc{} is steeper. At very low temperatures (see inset to Fig.~\ref{fig:thermopower}) \scc{} appears to change sign, although this should be confirmed by a more sensitive low-temperature technique.

Above 300\,K, \sab{} reaches a maximum of \seebeck{28} at around 400\,K and then declines gradually up to the highest measured temperature of 780\,K. A plateau in the value of $S$ at around \seebeck{30} is a common feature of many ruthenium oxides \cite{Hebert2015}. The data in this temperature range is compatible with a measurement on sintered polycrystalline material \cite{Keawprak2008}. 

It is only above 300\,K where \scc{} becomes considerably greater in absolute terms than \sab{}, reaching nearly \seebeck{50} at 600\,K before levelling off at higher temperatures. The structure of \scc{}$(T)$ is clearly of interest. The out-of-plane resistivity, $\rho_c$, peaks at $\sim 150$ K. This has been interpreted as a sign of the crossover between coherent transport at lower temperatures and incoherent transport at higher temperatures \cite{Hussey1998}. A point of inflection in the temperature dependence of \scc{} occurs at a similar temperature. 

This feature is unlikely to be related to phonon drag, as this generally requires that phonon transport in the same direction be out-of-equilibrium. This condition is usually only seen in the clean limit where the momentum-conserving electron-phonon scattering rate can outweigh the momentum-relaxing processes such as defect, impurity and umklapp scattering. While this might be achieved at low temperatures in \srtwo{} it is unlikely to be the case at 150\,K. The lack of a definitive phonon drag signature even in the low-resistance in-plane direction (typically a strong enhancement of $S$ at some fraction of the Debye temperature, with a dependence on sample purity) argues against phonon drag as an active mechanism.

\textit{Entropic interpretations}
In the low temperature limit, the slope of the Seebeck coefficient has been shown to track the electronic part of the specific heat with a proportionality fixed only by fundamental constants \cite{Behnia2004}. 
From the measured value of the electronic specific heat in the normal state \cite{Nishizaki2000}, we would expect $|S|/T \approx \sovert{0.4}$ as $T\rightarrow 0$. Comparison to the inset of Fig.\ref{fig:thermopower} shows that the data is of the correct order of magnitude, but that the low temperature limit has not been reached. 
Generally, in multi-band strongly correlated materials, the expected relationship is only seen at temperatures below 1K \cite{Behnia2004}.






At high temperature, the Heikes formula is derived in the atomic limit where transport proceeds by hopping between isolated sites \cite{Chaikin1976}. The Seebeck coefficient is given by $S_H = -(1/e)(\frac{\partial S_e}{\partial n}|_E)$ where $S_e$ is the entropy and $n$ is the carrier density. 
An alternative to this is the very similar Kelvin formula, $S_K = -(1/e)(\frac{\partial S_e}{\partial n}|_T)$, which is derived by taking transport formulae to the thermodynamic limit \cite{Peterson2010}.
In each case the Seebeck coefficient claims to be a measure of the entropy associated with each particle. 

Numerical evaluation of $S_H$ and $S_K$ for \srtwo{} \cite{Mravlje2016} produce values that are not too far away from the experimental ones at high temperature, at least for \sab{}. The continued rise of \scc{} with $T$ is not, however, captured by these calculations.

The other relevant implication of this interpretation is that $S$ should become isotropic in the limit of both high and low temperature if it is purely a reflection of the entropy transported by a carrier. However, as is seen here and in other materials such as BEDT-TTF \cite{Yu1991} and Nd-LSCO \cite{Gourgout2022}, anisotropy persists over a large temperature range.

\textit{Semi-classical transport} The transport coefficients in metallic materials generally depend on the scattering and dispersion of quasiparticles in an energy window of order $\pm 5k_B T$ around $\varepsilon_F$. Here, \scc{} only becomes very different from \sab{} for $T>300$K. If the assumptions of semi-classical transport hold at these high temperatures, this is an indication that it is states further than $\pm 0.1$ eV from $\varepsilon_F$ which contribute to changes in the anisotropy. \sab{} calculated from the electronic band structure \cite{Mravlje2016} has a quasi-linear temperature dependence that does not at all resemble the data. There is a significant peak in the out-of-plane transport function, but it is 1 eV below $\varepsilon_F$ and could therefore not be the origin of any feature in \scc{} observed here.


Some recent work has focused on the impact of the van Hove singularity in the $\gamma$ band on transport \cite{Zingl2019,Herman2019,Yang2023}.
At only 14meV above the Fermi level, the associated peak in the density of states should impact the Seebeck coefficient at lower temperatures than the apparent onset of anisotropy. It might therefore be tempting to assign the intermediate-temperature features in \scc{} to the presence of this van Hove singularity. However, the lack of associated features in \sab{} is difficult to understand without invoking an additional directional mechanism.

\textit{Mott formula} The Seebeck coefficient is expected to be relatively isotropic if it is governed by a single energy scale, even if the electrical conductivity is anisotropic. This can be understood by reference to the Mott formula, which relates the Seebeck coefficient to the electrical conductivity:

\begin{equation}
 S_i = \frac{\pi^2}{3}\frac{k_B}{e}T\frac{\partial(\ln \sigma_i)}{\partial \varepsilon}\bigg |_{\varepsilon = \varepsilon_F}
\end{equation}

\begin{figure}
\includegraphics[width=0.48\textwidth]{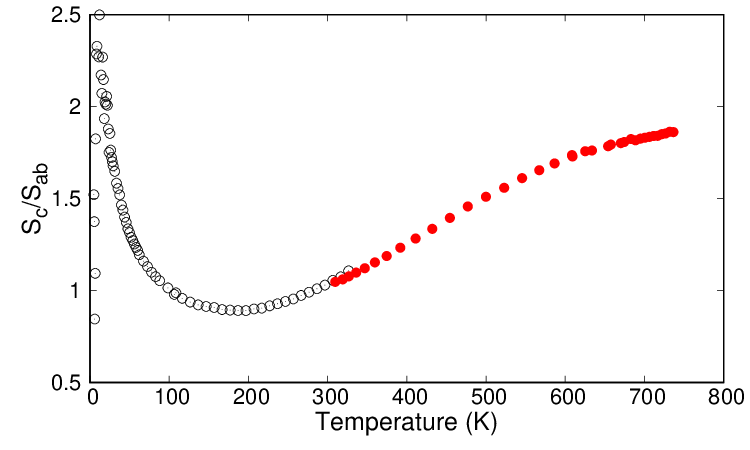}
\caption{Anisotropy of the Seebeck coefficient, $S_c/S_{ab}$, as a function of temperature. From around 100-300K it is close to unity and becomes larger below and above.}
\label{fig:anisotropy}
\end{figure}

The logarithmic derivative means that the Seebeck coefficient is much less sensitive to anisotropy in the electronic band structure than the conductivity. Even modest anisotropy in $S$ could therefore be an indication of a significant difference in the energy scale governing scattering in the two directions. The anisotropy $S_c/S_{ab}$ is shown in Figure \ref{fig:anisotropy}. While the low temperature behaviour (up to 100K) can be reasonably assigned to anisotropy originating in the electronic structure, it is by no means clear that the picture of semiclassical transport is still valid as the temperature increases further and the mean free path drops to the order of the lattice spacing. It is difficult therefore to explain the anisotropy in $S$ at high temperature using the Mott relation.

In one quasi-2D layered material, \pdcoo{}, a thermoelectric anisotropy of $\sim 50$ has been predicted to arise purely from band effects \cite{Ong2010}, but this has not been experimentally verified. The crucial factor behind this is the strong variation of the warping in the $k_z$ direction of the quasi-cylindrical Fermi surfaces as a function of energy. More typical is the case of intercalated graphite, where the thermoelectric power is anisotropic, but only by a factor of two, while the conductivity anisotropy is around 1000 \cite{Elzinga1982,Uher1985}.

In \srtwo{}, La-substitution was shown to donate electrons and induce a rigid shift of the electronic bands with respect to the Fermi energy, while introducing minimal disorder \cite{Kikugawa2004a}. The anisotropy in conductivity changes little \cite{Kikugawa2004b}, however, and we therefore expect $S$ to be isotropic at low temperatures, assuming that the changes in conductivity arise purely from the rigid shifting of the bands.


\textit{DMFT calculations} DMFT was used to calculate the transport coefficients of \srtwo{} in Ref.~\onlinecite{Mravlje2016}. A peak in the kernel of the out-of-plane transport integral was found at an energy $\sim 0.3$eV below the Fermi energy, causing an enhancement of \scc{} at temperatures approaching $\gtrsim 300$K.

While \scc{} is found to be greater than \sab{} everywhere, it is above 400 K that the calculated values begin to diverge considerably. The results of this calculation are reproduced in Fig.~\ref{fig:thermopower}. The significant trend in the experimental data is followed well. 

Transport calculations using the semi-classical model failed to reproduce either the rapid initial rise of \sab{} or the subsequent plateau. The semi-quantitative agreement of the data with DMFT shows that electron-electron correlations must be taken into account. The implication is that $S$ cannot generally be treated as a thermodynamic quantity, although $S_H$ and $S_K$ can be useful guides to the expected order of magnitude in the high temperature limit.

A similar conclusion was drawn from a recent exploration of the temperature dependence of the Hall effect \cite{Zingl2019}, which argued against the fine tuning that multi-band semi-classical transport calculations require to reproduce experimental data.

\textit{Comparison to other materials} The organic superconductor \salt{} displays a marked ansiotropy of $S$ within the conducting plane, as a result of quasi-1D features of the electronic structure\cite{Yu1991}. While a tight-binding model was able to reproduce the signs and temperature dependence of $S$ up to 300K quite well, the bandwidth had to be artifically reduced to obtain this agreement, suggesting that strong correlations had to be taken into account. Like \srtwo{}, \salt{} has a relatively low coherence temperature of around 50K. DMFT calculations \cite{Merino2000} later treated the crossover into the incoherent regime and were able to reproduce the features seen in $S$.

The high temperature cuprate superconductors are also incoherent, anisotropic metals in their normal state. In \lsco{} the temperature evolution of the Seebeck coefficient measured within and perpendicular to the CuO$_2$ planes is very reminiscient of \srtwo{} \cite{Nakamura1993}. At $x=0.10$, the magnitudes and anisotropy are similar to that seen in \srtwo{}, with \sab{} rising to a broad maximum, while \scc{} continues to rise up to 300 K. On further doping, \sab{} drops and changes sign while \scc{} retains similar features.

A recent study on Nd-\lsco{} \cite{Gourgout2022}, where superconductivity and its associated impact on transport is suppressed to low temperatures, also revealed strong anisotropy across a wide range of doping. A particular feature is the appearance of non-monotonic features in \scc{}(T) at around 100\,K, which are not seen in \sab{}. This parallel with \srtwo{} merits further investigation.

\textit{Conclusions} 
We observe a strong anisotropy in the Seebeck coefficient of \srtwo{} at temperatures above 300K, in accordance with predictions of transport calculations. The out-of-plane Seebeck coefficient contains features at an intermediate temperature similar to that of the peak in c-axis resistivity which merit further consideration. The isotropic behaviour predicted by arguments based on the entropy of the carriers is not realized, neither at the moderately high temperatures where these interpretations have enjoyed some success \cite{Marsh1996,Berggold2005}, nor at temperatures down to 5K. While ab initio calculations do appear to provide some insight, these results highlight the lack of an intuitive framework to understand the anisotropy of thermopower at high temperatures.



\begin{acknowledgments}
The authors would like to thank J. Mravlje, A. Georges and K. Behnia for stimulating discussions, and P. Kushwaha for helping to source the crystal used for the measurements of the in-plane thermopower. YM is supported by JSPS Core-to-core program JPJSCCA20170002 and JSPS Kakenhi JP22H01168.
\end{acknowledgments}

%

\end{document}